\documentclass[twocolumn,aps,prl,superscriptaddress,showpacs,secnumroman,showkeys]{revtex4}
\usepackage{color}%
\usepackage{amsmath,bm}%
\usepackage{graphicx}%
\begin{document}
\title{A novel determination of density, temperature and symmetry energy for nuclear multi-fragmentation through primary fragment yield reconstruction}
\author{W. Lin}
\affiliation{Institute of Modern Physics, Chinese Academy of Sciences, Lanzhou, 730000, China}
\affiliation{University of Chinese Academy of Sciences, Beijing 100049, China}
\author{X. Liu}
\affiliation{Institute of Modern Physics, Chinese Academy of Sciences, Lanzhou, 730000, China}
\affiliation{University of Chinese Academy of Sciences, Beijing 100049, China}
\author{M. R. D. Rodrigues}
\affiliation{Instituto de F\'{\i}sica, Universidade de S\~{a}o Paulo, Caixa Postal 66318, CEP 05389-970, S\~{a}o Paulo, SP, Brazil}
\author{S. Kowalski}
\affiliation{Institute of Physics, Silesia University, Katowice, Poland.}
\author{R. Wada}
\email[E-mail at:]{wada@comp.tamu.edu}
\affiliation{Institute of Modern Physics, Chinese Academy of Sciences, Lanzhou, 730000, China}
\author{M. Huang}
\affiliation{Institute of Modern Physics, Chinese Academy of Sciences, Lanzhou, 730000, China}
\author{S. Zhang}
\affiliation{Institute of Modern Physics, Chinese Academy of Sciences, Lanzhou, 730000, China}
\affiliation{University of Chinese Academy of Sciences, Beijing 100049, China}
\author{Z. Chen}
\affiliation{Institute of Modern Physics, Chinese Academy of Sciences, Lanzhou, 730000, China}
\author{J. Wang}
\affiliation{Institute of Modern Physics, Chinese Academy of Sciences, Lanzhou, 730000, China}
\author{G. Q. Xiao}
\affiliation{Institute of Modern Physics, Chinese Academy of Sciences, Lanzhou, 730000, China}
\author{R. Han}
\affiliation{Institute of Modern Physics, Chinese Academy of Sciences, Lanzhou, 730000, China}
\author{Z. Jin}
\affiliation{Institute of Modern Physics, Chinese Academy of Sciences, Lanzhou, 730000, China}
\affiliation{University of Chinese Academy of Sciences, Beijing 100049, China}
\author{J. Liu}
\affiliation{Institute of Modern Physics, Chinese Academy of Sciences, Lanzhou, 730000, China}
\author{F. Shi}
\affiliation{Institute of Modern Physics, Chinese Academy of Sciences, Lanzhou, 730000, China}
\author{T. Keutgen}
\affiliation{FNRS and IPN, Universit\'e Catholique de Louvain, B-1348 Louvain-Neuve, Belgium}
\author{K. Hagel}
\affiliation{Cyclotron Institute, Texas A$\&$M University, College Station, Texas 77843}
\author{M. Barbui}
\affiliation{Cyclotron Institute, Texas A$\&$M University, College Station, Texas 77843}
\author{C. Bottosso}
\affiliation{Cyclotron Institute, Texas A$\&$M University, College Station, Texas 77843}
\author{A. Bonasera}
\affiliation{Cyclotron Institute, Texas A$\&$M University, College Station, Texas 77843}
\affiliation{Laboratori Nazionali del Sud, INFN,via Santa Sofia, 62, 95123 Catania, Italy}
\author{J. B. Natowitz}
\affiliation{Cyclotron Institute, Texas A$\&$M University, College Station, Texas 77843}
\author{E. J. Kim}
\affiliation{Cyclotron Institute, Texas A$\&$M University, College Station, Texas 77843}
\affiliation{Division of Science Education, Chonbuk National University, Jeonju 561-756, Korea}
\author{T. Materna}
\affiliation{Cyclotron Institute, Texas A$\&$M University, College Station, Texas 77843}
\author{L. Qin}
\affiliation{Cyclotron Institute, Texas A$\&$M University, College Station, Texas 77843}
\author{P. K. Sahu}
\affiliation{Cyclotron Institute, Texas A$\&$M University, College Station, Texas 77843}
\author{K. J. Schmidt}
\affiliation{Institute of Physics, Silesia University, Katowice, Poland.}
\affiliation{Cyclotron Institute, Texas A$\&$M University, College Station, Texas 77843}
\author{S. Wuenschel}
\affiliation{Cyclotron Institute, Texas A$\&$M University, College Station, Texas 77843}
\author{H. Zheng}
\affiliation{Cyclotron Institute, Texas A$\&$M University, College Station, Texas 77843}
\affiliation{Physics Department, Texas A$\&$M University, College Station, Texas 77843}

\date{\today}

\begin{abstract}
For the first time primary hot isotope distributions are
experimentally reconstructed in intermediate heavy ion collisions and
used with antisymmetrized molecular dynamics (AMD) calculations to
determine density, temperature and symmetry energy coefficient in a
self-consistent manner.
A kinematical focusing method is employed to reconstruct the primary hot fragment yield distributions for multifragmentation events observed in the reaction system $^{64}$Zn + $^{112}$Sn at 40 MeV/nucleon.
The reconstructed yield distributions are in good agreement with the primary isotope distributions of AMD simulations.
The experimentally extracted values of the symmetry energy coefficient relative to the temperature, $a_{sym}/T$, are compared with those of the AMD simulations
with different density dependence of the symmetry energy term.
The calculated $a_{sym}/T$ values changes according to the different interactions. By comparison of the experimental values of $a_{sym}/T$ with those of calculations, the density of the source at fragment formation was determined to be $\rho /\rho_{0} = (0.63 \pm 0.03 )$. Using this density, the symmetry energy coefficient and the temperature are determined in a self-consistent manner as $a_{sym} = (24.7 \pm 1.9) MeV$ and $T=(4.9 \pm 0.2)$ MeV.

\end{abstract}
\pacs{25.70Pq}

\keywords{Intermediate Heavy ion reactions, reconstructed multiplicity of primary isotopes, kinematical focusing method, secondary statistical decay, density, temperature and symmetry energy}

\maketitle


Constraining the density dependence of the symmetry energy is one of
the key objectives of contemporary nuclear physics. It plays a key
role for various phenomena in nuclear astrophysics, nuclear
structure, and nuclear reactions ~\cite{Lattimer04,BALi08}. Heavy ion
collisions provide a unique opportunity to study the nuclear symmetry
energy and its density dependence at and around normal nuclear matter
density. However, reliable extraction is difficult because of the
complexity of the reaction dynamics.

In violent heavy ion collisions in the intermediate energy regime (20 $\leq E_{inc} \leq$ a few hundred MeV/nucleon), intermediate mass fragments (IMFs) are copiously produced through a multi-fragmentation process.  Nuclear multi-fragmentation was predicted a long time ago~\cite{Bohr36} and has been studied extensively following the advent of 4$\pi$ detectors. Studies of nuclear multi-fragmentation provide important information on the properties of the hot nuclear matter equation of state. The recent status of the experimental and theoretical work is reviewed in Refs.~\cite{Borderie08,Gulminelli06,Chomaz04}.

In general the nuclear multi-fragmentation process, can be divided into three stages, i.e., dynamical compression and heating, expansion and freeze-out of  primary fragments, and finally the separation and cooling of the primary fragments by evaporation.

Different models have been developed to model the multi-fragmentation process. These include dynamical transport models such as
FMD~\cite{Feldmeier90}, AMD~\cite{Ono96,Ono99,Ono02}
CoMD~\cite{Papa01}, ImQMD~\cite{Wang02}, QMD~\cite{Aichelin91}, BNV~\cite{Kruse85}, SMF~\cite{Colonna98}, BUU~\cite{Aichelin85} among others. Most of these can account reasonably well for many characteristic properties experimentally observed. On the other hand statistical multi-fragmentation models such as MMMC~\cite{Gross90} and SMM~\cite{Bondorf85},
based on a quite different assumption from the transport models, can also describe many experimental observables well. The statistical models assume that fragment formation takes place at freeze-out in equilibrated low density nuclear matter. Typically the source parameters such as size, neutron/proton ratio, density and temperature are optimized to reproduce the final state experimental observables.  In contrast, the transport models do not assume any chemical or thermal equilibration a priori.
When fragments are initially formed in the multi-fragmentation process, many of them are in excited states~\cite{Marie98,Hudan03,Rodrigues13}. These "primary hot" fragments will de-excite by evaporation processes before they are detected as "secondary cold" fragments. This is also true in the statistical multifragmentation models. This cooling process may significantly alter the fragment yield distributions of the observed cold isotopes from the primary ones which directly reflect the density and temperature of the fragmenting source~\cite{Huang10_1,Huang10_2,Huang10_4,Chen10,Bonasera08}. Even though the statistical decay process itself is rather well understood and well coded, it is not a trivial task to combine it with a dynamical code. The statistical evaporation codes assume the fragments to be at thermal equilibrium with normal nuclear densities and shapes. These conditions are not guaranteed for fragments when they are formed in the multifragmentation process.

In order to avoid the complication of secondary decay modification of fragment yields we previously proposed experimental methods for kinematical reconstruction of the primary fragment yields and excitation energies in complex multifragmentation events~\cite{Marie98,Hudan03,Rodrigues13}.
In this letter we report on the use of experimental reconstruction of primary fragment yields to characterize the fragmenting source, using the ratio, $a_{sym}/T$. In a transport model such as AMD, the dynamic evolution of the system is such that variations in the temperature and density and symmetry energy, are closely correlated with each other. If one of these parameter is determined, then other parameters can be extracted in a self-consistent manner from the transport model solutions using these relationships. In the analysis presented here, the density of the fragmenting source is determined first and the temperature and symmetry energy are extracted using the model predicted correlations.
This is the first time such a self-consistent analysis has been performed on experimental primary fragment data.
A further detailed description of the analysis presented here will be given in a forthcoming article~\cite{Lin2014}.



The experiment was performed at the K-500 superconducting cyclotron facility at Texas $A\&M$ University. $^{64,70}$Zn and $^{64}$Ni beams were used to irradiate $^{58,64}$Ni, $^{112,124}$Sn, $^{197}$Au, and $^{232}$Th targets at 40 MeV/nucleon. This article focuses on the $^{64}$Zn + $^{112}$Sn reaction. Details of the experiment have been given in Refs.~\cite{Rodrigues13}. Here we briefly outline the experiment. Intermediate mass fragments (IMFs; 3 $\leq$ Z $\leq$ 18) were detected by a detector telescope placed at $\theta_{lab} = 20^{o}$. This telescope provided the main trigger for all detected events. As discussed in detail in Ref.~\cite{Huang10_2}, the events measured by this IMF trigger belong essentially to the event class of semi-central collisions. Six to eight isotopes for $3 \leq Z \leq 18$ were typically clearly identified. Two sets of detectors were used to detect the light particles (LPs). For the light charged particles (LCPs), 16 single-crystal CsI(Tl) detectors of 3 cm length were set around the target at the opening angles between the trigger telescope and the detector $17^{o} \le \theta_{IMF-p} \le 155^{o}$, tilted $30^{o}$ in the azimuthal angle to avoid shadowing the neutron detectors described below.
The pulse shape discrimination method was used to identify p, d, t , $^{3}He$ and $\alpha$ particles. For neutrons 16 detectors of the Belgian-French neutron detector array, DEMON(DEtecteur MOdulaire de Neutrons)~\cite{Tilquin95}, were used. The detectors were distributed to achieve opening angles between the telescope and the detector of $15^{o} \le \theta_{IMF-n} \le 160^{o}$.


A kinematical focusing technique was employed to evaluate the LP multiplicities associated with each isotopically identified IMF. A detailed description of the kinematical focusing analysis is given in Ref.~\cite{Rodrigues13}. In Fermi energy heavy ion collisions, light particles are emitted at different stages of the reaction and from different sources during the evolution of the collision. The majority of light particles emitted in a violent event are uncorrelated ones and therefore it is crucial to distinguish the particles emitted from precursors of a particular IMF from the uncorrelated particles emitted from other sources. In the following the particles emitted from a precursor IMF are designated "correlated" particles and those not emitted from the precursor IMF are designated as "uncorrelated" particles. The separation of these contributions is based upon a kinematical focusing analysis: When correlated particles are emitted from a moving parent of an IMF, whose velocity $v_{IMF}$ is approximated by the velocity of the detected trigger IMF, the particles isotropically emitted in the frame of the IMF tend to be kinematically focused into a cone centered along the $v_{IMF}$ vector of the detected IMF. This is not the case for uncorrelated particles emitted in the same event. In our analysis
the contribution of the correlated particles was determined by use of a moving source parametrization and the shape of the uncorrelated spectrum was obtained from the particle velocity spectrum observed in coincidence with Li isotopes which are accompanied by the least number of  correlated  particles. As a result correlated particle multiplicities extracted for a given isotope need to be corrected by the addition of an amount corresponding to the correlated emission of that particle from the Li isotopes.
The amount added was evaluated from the AMD-GEMINI simulations~\cite{Ono99,Charity88}. The correlated  multiplicities were extracted for n, p, d, t and $\alpha$ particles. The correlated $^{3}He$ yields were very small and it was not possible to extract $^{3}He$ yields. Therefore the $^{3}He$ contribution was neglected in this work. See Ref.~\cite{Rodrigues13} for further detail.

Since only the average values of LP multiplicities can be extracted from this experiment, the widths of the multiplicity distributions have been evaluated, using the statistical decay code, GEMINI~\cite{Charity88}. A detailed explanation of the method is given in Ref.~\cite{Rodrigues13}. The widths of the multiplicity distributions of light particles associated with all experimentally observed final fragments were evaluated at the primary fragment excitation energies of $2.25\pm 0.25$ MeV/nucleon, which was previously found to be the average excitation energy of the primary fragments.

The multiplicities of the primary isotopes were then reconstructed using a Monte Carlo method~\cite{Lin2013}.
For the reconstruction, LP multiplicities, $M_{i}(i=n,p,d,t,\alpha)$, are generated for a given cold daughter nucleus on an event by event basis, assuming Gaussian distributions with a width evaluated by the GEMINI simulation, and their centroid is adjusted to give the same average multiplicity as that of the experiment. Then the mass and charge of the primary isotope, $A_{hot}, Z_{hot}$ was calculated as $A_{hot}=\sum_{i} M_{i} A_{i} + A_{cold}$ and $Z_{hot}=\sum_{i} M_{i} Z_{i} + Z_{cold}$, where
$A_{i}, Z_{i}$ are the mass and charge of correlated particle {\it i} and $A_{cold}, Z_{cold}$ are those of the detected isotope.

The Monte Carlo code was used to generate 100,000 parents for each detected IMF. From these results the primary fragment yield distributions were determined using the experimental multiplicity as a weighting factor. The multiplicities associated with the unstable nuclei of $^{8}Be$ and $^{9}B$ were added artificially by estimating their multiplicity and associated LP multiplicities from the neighboring isotopes.

\begin{figure}[htb]
\includegraphics[scale=0.4]{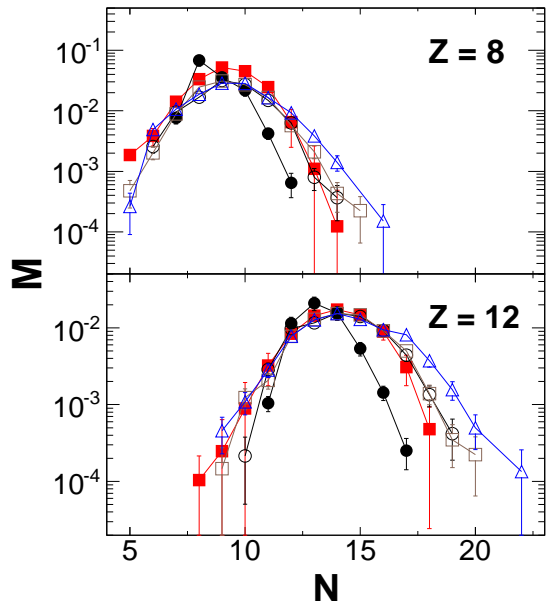}
\caption{\footnotesize
(Color online) Typical isotopic multiplicity distributions of experimental cold (closed circles), reconstructed primary (closed squares) as well as AMD primary fragments with g0 (open triangles), g0AS (open squares) and g0ASS (open circles) as a function of fragments neutron number $N$ for a given charge Z, which is indicated in the figure.
}			
\label{fig:Isotopic_distribution}
\end{figure}

In Fig.\ref{fig:Isotopic_distribution} the typical yield distributions of the reconstructed hot isotopes are compared with those of the experimentally observed cold isotopes for Z=8 and 12. These are representative distributions among those of the other elements. The reconstructed yields (closed squares) show much wider distributions than those of the cold isotopes (dots), which indicates the significant modification of the yield distributions between the primary hot and the observed cold isotopes, caused by the secondary decay process. The reconstructed yield distributions are further compared with those for AMD primary fragments observed at $t = 300 fm/c$ and calculated using three different Gogny interactions, the standard Gogny interaction (g0), an asymptotic stiff interaction (g0AS) and an asymptotic super-stiff interaction (g0ASS), having different density dependencies of the symmetry energy term ~\cite{Ono99,Ono03}. The comparisons are made in absolute multiplicity.
The reconstructed primary isotopic distributions are reasonably well reproduced by those of the AMD simulations with three different interactions. Different choices of the density dependence of the symmetry energy term give notable differences for the very neutron or proton rich isotopes. The variance of the distributions becomes maximum for g0 and minimum for g0ASS in most of the cases. The experimentally reconstructed yield  distributions favor the minimum variance distributions of g0ASS more quantitative comparisons are made below.

In order to extract the characteristics of the emitting source of these isotopes, we first determined the ratio of the symmetry energy coefficient relative to the temperature, $a_{sym}/T$, employing the  the isobaric yield ratio method~\cite{Huang10_1,Huang10_2,Chen10,Huang10_3}.
The isobaric yield method is based on the Modified Fisher Model (MFM)\cite{Fisher1967}, which has been used to study the properties of the hot nuclear matter in previous work~\cite{Huang10_1}. According to MFM, $a_{sym}/T$ can be extracted using the yield ratio of two isobars in single reaction system as~\cite{Huang10_1},
\begin{eqnarray}
a_{sym}/T &= -\frac{A}{8}\{\ln[R(3, 1, A)]-\ln[R(1, -1, A)] \nonumber\\
&-\Delta(3, 1, A) + \Delta E_{c}\}
\label{eq:Symmetry}
\end{eqnarray}
where $R(1, -1, A)= Y(1,A)/Y(-1,A)$ and $Y(I,A)$ is the yield of isotope with $I=N-Z$ and A. $\Delta(3, 1, A)$ is the difference in mixing entropies of isobars $A$ with $I = 3$ and 1. $\Delta E_{c}$ is the difference of the Coulomb energy between the neighboring isobars and given by $\Delta E_{c} = 2a_{c}/(A^{1/3}T)$. One should note that the values of $\Delta(3, 1, A)$ and $\Delta E_{c}$ are small compared to the first two terms and they have opposite signs each other. $a_{c}/T$ values are taken from Ref.~\cite{Huang10_1}.

\begin{figure}[htb]
\includegraphics[scale=0.4]{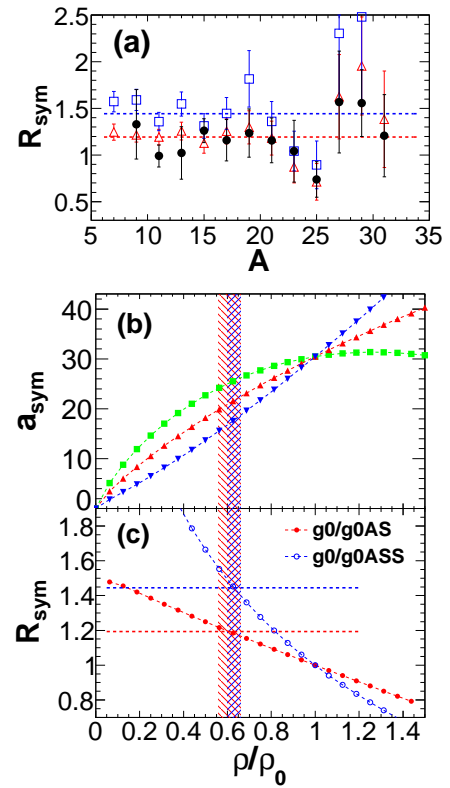}
\caption{\footnotesize
(Color online) (a) Ratios of the calculated $a_{sym}/T$ values for g0/g0AS (open squares), g0/g0ASS (open triangles) and g0/the reconstructed experimental yield (dots). Dotted lines are the average values for the AMD simulations. The values are given in Table~\ref{table:data_results}. (b) Symmetry energy coefficient vs density for g0(squares), g0AS(upward triangles) and g0ASS (downward triangles) used in the AMD simulations. The shaded vertical area indicates the fragment formation density extracted from the ratio of the symmetry energy coefficient. (c) Ratio of the symmetry energy coefficient in (b) between g0AS/g0 and g0ASS/g0 as a function of the density. The horizontal dotted lines indicates the ratio values extracted from the $a_{sym}/T$ values in (a).
}			

\label{fig:Symmetry_AMD}
\end{figure}

Values of $a_{sym}/T$ were extracted from the experiment and from the results of AMD simulations with g0, g0AS, and g0ASS interactions and some of them were previously discussed in Ref.~\cite{Huang10_1}.
Primary isotope values obtained from the AMD simulations are almost constant below $A = 25$ and those for the three interactions - g0, g0AS and g0ASS - are  more or less parallel to each other. The ratios $a_{sym}/T$ for g0 relative to those for g0AS and g0ASS are plotted in Fig.\ref{fig:Symmetry_AMD}(a), together with the ratio of those from g0 relative to those from the reconstructed yields (dots). Both of the calculated ratios are more or less constant as a function of $A$, though those from the reconstructed yields distribute around the values of g0/g0ASS values with a slightly larger fluctuation than those of the simulations.
Following Ref.~\cite{Ono03}, we interpret the ratios as resulting from the difference of the symmetry energy coefficient at a given density and temperature at the fragment formation. In Fig.\ref{fig:Symmetry_AMD}(b), the density dependence of the symmetry energy coefficient for g0, g0AS and g0ASS is shown as a function of $\rho / \rho_{0}$. In Fig.\ref{fig:Symmetry_AMD}(c), the  ratios for g0/g0AS and g0/g0ASS are shown. The two horizontal dotted lines in Fig.\ref{fig:Symmetry_AMD}(c) show the ratios extracted from Fig.\ref{fig:Symmetry_AMD}(a) and the values are given in the first column of Table~\ref{table:data_results}.
From these ratio values the densities are extracted as indicated by the vertical shade areas in Fig.\ref{fig:Symmetry_AMD}(b) and (c) and given in the second column of Table~\ref{table:data_results}. Assuming the nucleon density is same for the three different interactions used, the nucleon density of the fragmenting system is determined from the overlap value of the extracted values. This assumption is reasonable because the nucleon density is mainly determined by the stiffness of the EOS and not by the density dependence of the symmetry energy term.
$\rho / \rho_{0} = 0.63 \pm 0.03$ is obtained.
One should note that the errors on the density values evaluated here and those of the $a_{sym}/T$ values in the fourth column govern the errors on the temperature and symmetry energy extracted below since they are determined using their predicted correlations in the AMD model.

The corresponding symmetry energy coefficient values of the calculations are extracted from Fig.\ref{fig:Symmetry_AMD}(b) and given in the third column.
In the fourth column, the average $a_{sym}/T$ values from the calculations and the reconstructed isotopes are shown. For the AMD simulations, the temperature is calculated by $T=a_{sym}/(a_{sym}/T)$ and given in the fifth column. From the temperature values for the AMD simulations with different interactions, the overlapping values of $T = 4.9 \pm 0.2 MeV$ is assigned to the temperature at the time of the fragment formation. Using this temperature and the experimental $a_{sym}/T$ value in the bottom of the fourth column, the experimental symmetry energy coefficient is determined as $a_{sym} = 24.7 \pm 1.9 MeV$.
These extracted symmetry energy coefficient, temperature and density for the fragment formation show notable differences from those of Ref.~\cite{Shetty07}, where values were extracted from the experimentally observed secondary yields using isoscaling parameters. In their work, reactions of $^{40}$Ar, $^{40}$Ca + $^{58}$Ni,$^{58}$Fe at 25-55 MeV/nucleon were studied.
In Table~\ref{table:data_results}, the values at the excitation energy of $E^{*} = 7 MeV$ are given for a comparison.

\begin{table}[ht]
\caption{Extracted parameters. The values from Ref.~\cite{Shetty07} are taken at $E^{*} = 7 MeV$ and they depend slightly on the excitation energy.} 
\centering 
\begin{tabular}{c c c c c c} 
\hline\hline 
 & Ratio & $\rho/\rho_{0}$ & $a_{sym}$ & $a_{sym}/T$ & T\\ [0.5ex] 
 &       &                 & (MeV)     &             & (MeV)\\ [0.5ex] 
\hline 
g0    &               &               & 25.7$\pm$0.6 & 5.29$\pm$0.13 & 4.9$\pm$0.2\\
g0AS  & 1.19$\pm$0.03 & 0.61$\pm$0.05 & 21.2$\pm$1.2 & 4.31$\pm$0.12 & 4.9$\pm$0.4\\
g0ASS & 1.44$\pm$0.05 & 0.63$\pm$0.03 & 17.8$\pm$0.9 & 3.50$\pm$0.12 & 5.1$\pm$0.5\\
\hline 
Exp   &               & 0.63$\pm$0.03 & 24.7$\pm$1.9 & 5.04$\pm$0.32 & 4.9$\pm$0.2\\[0.5ex]
Ref.~\cite{Shetty07} &    & 0.45$\pm$0.12 & 17$\pm$2  &              & 6.5$\pm$0.5\\[0.5ex]
\hline 
\end{tabular}
\label{table:data_results} 
\end{table}

Summarizing, the yield distribution of primary hot isotopes has been reconstructed experimentally for the $^{64}$Zn + $^{112}$Sn reaction at 40 MeV/nucleon, employing a kinematical focusing technique.
The reconstructed primary isotope multiplicities are in good agreement with those calculated in AMD simulations. From the reconstructed distributions, the ratio $a_{sym}/T$ as a function of $A$ was evaluated using the isobaric yield ratio method.
Employing the density determined from AMD simulations, $\rho / \rho_{0} = 0.63 \pm 0.03 $, a temperature of $T = 4.9 \pm 0.2 MeV$ and the symmetry energy coefficient of $a_{sym} = 24.7 \pm 1.9 MeV$ are extracted in a self-consistent manner for the fragmenting source of the experimentally reconstructed primary yields.

\section*{Acknowledgments}

We thank the staff of the Texas A$\&$M Cyclotron facility for their support during the experiment. We thank the Institute of Nuclear Physics of the University of Louvain and Prof. Y. El Masri for allowing us to use the DEMON detectors. We thank A. Ono and R. J. Charity for providing their codes. This work is supported by the U.S. Department of Energy under Grant No.DE-FG03-93ER40773, the Robert A. Welch Foundation under Grant A0330, the National Natural Science Foundation of China (Grants No. 11075189), 100 Persons Project (0910020BR0, Y010110BR0), ADS project 302 (Y103010ADS) of the Chinese Academy of Sciences and LG Yonam Foundation 2012 in Korea.









\end{document}